\documentclass[12pt,twocolumn]{article}
\usepackage{widetext}
\usepackage{flushend}
\usepackage{cuted}
\usepackage{color}
\usepackage{graphicx}
\usepackage{hyperref}
\usepackage{pstricks,amssymb}
\usepackage{fancyhdr}
\usepackage[raggedright]{titlesec}
\begin{document}
\title{The Paradox of Two Charged Capacitors -- A New Perspective\\
\vspace{.3cm}
\large{Ashok K. Singal}\\
\vspace{0.3cm}
\normalsize{
Astronomy and Astrophysics Division\\
Physical Research Laboratory\\
Navrangpura, Ahmedabad - 380 009, India.\\
{\small asingal@prl.res.in}\\
\vspace{0.3cm}
(Submitted 10-07-2015)}\\
\rule[0.1cm]{16cm}{0.02cm} \\
\textbf{Abstract}\\
\flushleft \normalsize {It is shown that the famous paradox of two charged capacitors is successfully resolved 
if all the energy changes in the system are properly considered when some of the charges are 
transferred from one capacitor to the other. It happens so even when the connecting wire has 
an identically zero resistance, giving rise to no Ohmic losses in the wire. 
It is shown that in such a case the ``missing energy'' goes into the kinetic energy of 
conducting charges. It is shown that radiation plays no significant role in resolving 
the paradox. 
The problem can be formulated and successfully resolved in a novel form, 
where the capacitance of the system is increased by stretching the plates of the original capacitor, 
without involving any connecting wires in a circuit. There is an outward self-force due to mutual repulsion 
among charges stored within each capacitor plate, and the work done by these self-forces during an expansion 
is indeed equal to the missing energy of the capacitor system.
} \\
\rule[0.1cm]{16cm}{0.02cm} 
}
\date{}
\maketitle
\thispagestyle{fancy} 
\lhead{\textbf{Physics Education}}
\chead{\thepage}
\rhead{\bf {dateline}(to be added by Editor)}
\lfoot{Volume xx, Number y Article Number : n.(to be added by editor) }
\cfoot{ }
\rfoot{www.physedu.in}
\renewcommand{\headrulewidth}{0.4pt}
\renewcommand{\footrulewidth}{0.4pt}
\section{Introduction}
In the famous two-capacitor paradox\cite{1,6,7,10,11,12,13,14,15,16,18} one of the capacitors, 
say C$_{1}$, of capacitance $C$ is initially charged to a voltage $V_{0}$ with charge 
$Q_{0}=CV_{0}$ and energy $U_0=CV_{0}^2/2=Q_{0}V_{0}/2=Q_{0}^2/(2C)$, 
while the other similar capacitor, C$_{2}$, is initially uncharged, thereby the total 
energy of the system being $U_0$. Both capacitors are assumed be to identical in every 
respect. Now C$_{1}$ is connected to C$_{2}$ using a conducting wire, resulting in transfer of 
some charges from C$_{1}$ to C$_{2}$. From symmetry 
each capacitors will end up with charge $Q_{0}/2$ and voltage $V_{0}/2$, with energy of each
as $CV_{0}^2/8=U_0/4$. Therefore the total energy of the system will be $U_0/2$.
What happened to the other half of the energy?

Puzzling though this might appear at a first look, the loss of energy is easily explained 
if we consider the Ohmic losses in the connecting wires. Suppose the connecting wires have 
a resistance $R$ (Fig.~1), then the charging current will be $(V_{0}/R)e^{-2t/(RC)}$ and the
dissipated energy will be,
\begin{eqnarray}
\label{1}
\nonumber
\int^{\infty}_{o}I^2 R \:{\rm d}t&=\int^{\infty}_{o}\left(\frac{V_{0}}{R}e^{-2t/(RC)}\right)^2 R \:{\rm d}t\\
&= \frac{CV_{0}^{2}}{4}=\frac{U_0}{2}.
\end{eqnarray}
The above equation is true for any finite value of $R$. But what happens if  
there were no Ohmic losses, e.g., if in our ideal hypothetical case the 
resistance were identically zero (a superconductor!). The total energy in the two
capacitors, however, is still half of the initial energy, so where does the remaining 
energy disappear?
\begin{figure}[ht]
\scalebox{0.45}{\includegraphics{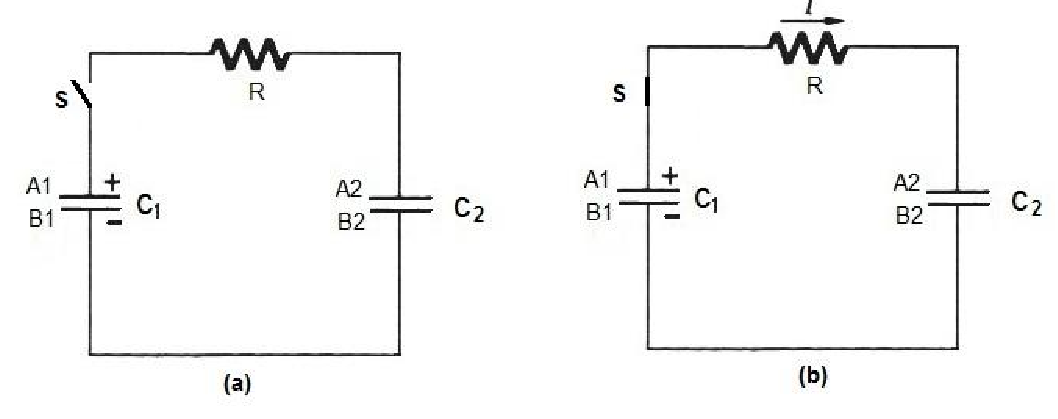}}
\caption{Charging a parallel plate capacitor.}
\end{figure}

Of course there is nothing special about the two capacitors being identical. In the case the two 
capacitances $C_1$ and $C_2$ are unequal, the initial stored energy $U_0=Q_{0}^2/2C_1$ after transfer 
of charges reduces to $Q_{0}^2/(2(C_1+C_2))=U_0 C_1/(C_1+C_2)$. This implies a loss of energy\cite{19} 
\begin{equation}
\label{1a}
\Delta U = \frac{U_0\; C_2}{C_1+C_2}. 
\end{equation}
For equal capacitances ($C_1=C_2$) the energy loss reduces to $U_0/2$, as derived 
earlier. Of particular interest is the case for large $C_2$ ($C_2\rightarrow \infty$), where  
all stored energy is lost. 

Since the charges undergo acceleration while moving from higher to a lower potential in case of zero resistance, 
can it be that whole of the missing energy appears as radiation from these accelearted charges?
The current belief seems to be that the missing energy is radiated away.\cite{8,9} It should be clarified that here we are not 
talking of the thermal electromagnetic radiation like in a resistance wire, but of 
electromagnetic waves radiated from an antenna system. 
As we will show in Section 4, the present radiation calculations are based on circular arguments. Moreover 
from maximum possible radiation losses from Larmor's formula we will argue that missing energy cannot be accounted for 
by radiation losses, and that the radiation hypothesis does not offer a satisfactory resolution of the paradox.

\section{Where does the missing energy go?}
The missing energy actually goes into the kinetic energy of conducting charges getting transferred 
from C$_1$ to C$_2$ for $R=0$\cite{19}.
Actually one has to be cautious when extremely low resistances are considered. The conductivity of a 
metal is directly proportional to the characteristic time $\tau$ between successive 
collisions of the charge carriers that results in loss of directional correlation\cite{1,3}. 
Drift velocity in the conductor is $qE\tau/m$, where $E$ is the electric field
and $q$ is the electric charge and $m$ is the mass of the  charge carrier (an electron!).
A typical value for $\tau$ in the metals is $\approx 10^{-14}$ sec  with typical drift 
velocity usually a fraction of a mm/sec. 
The resistivity is $\propto 1/\tau$, and low resistivity implies $\tau$ is large 
and then the mean free path $\lambda$ between collisions ($\propto \tau $) would also be large. In that 
case there will be fewer collisions and in an extreme case, we could assume that 
the mean free path $\lambda$ will be large enough to be longer than the length of the wire or channel joining the 
two capacitors. This could be termed as $R=0$ case.
Then the conducting charges will steadily gain velocity and kinetic energy as the 
collisions will be minimal. In that case the charges will not undergo Ohmic losses and when they 
reach C$_2$ their kinetic energy  will be equal to the potential energy difference during the transfer 
between two capacitors. 

The gain in kinetic energy in the absence of Ohmic losses is easily calculated from the change in 
potential energy of each charge. 
The charge gains a velocity increment $\Delta v=qE\Delta t/m$ or $m\Delta v=q E\Delta x/v$ which implies 
a kinetic energy gain $\Delta (m v^2/2)=q\Delta V$.
For a charge transfer $Q$ from C$_1$ to C$_2$, the voltage difference 
between the two becomes $\Delta V=(Q_0-Q)/C_1 - Q/C_2$. Then the total kinetic energy gained by charges during a total 
charge transfer $Q_2$ from C$_1$ to C$_2$ is,
\begin{eqnarray}
\label{2}
\nonumber
\int^{Q_2}_{0}\left(\frac{Q_0-Q}{C_1} - \frac{Q}{C_2}\right) \:{\rm d}Q=\frac{Q_0 Q_2}{C_1}\\
-\frac{Q^2_2}{2}\left(\frac{1}{C_1}+\frac{1}{C_2}\right).
\end{eqnarray}
As the voltage difference between C$_1$ to C$_2$ becomes zero at the end, then $(Q_0-Q_2)/C_1 = Q_2/C_2 = Q_0/(C_1 + C_2)$, 
implying that the total kinetic energy gained by the charges from (\ref{2}) is 
$Q_2 Q_0/(2C_1)=Q_2 V_0/2=U_0C_2/(C_1+C_2)$, in agreement with the energy loss $\Delta U$ in (\ref{1a}).

When the charges finally get deposited on plates of the capacitor C$_2$ this kinetic 
energy should get transferred to the plates of C$_2$, which as we discuss later, could even be utilized by an 
external agency, or else the plates of C$_2$ would get heated.
The problem as posed is between two equilibrium states in which the charges are stationary 
both initially and in the final state. Thus there should be no residual kinetic energy in the system. 
It implies the charges when finally get deposited on plates of the capacitor C$_2$, they remain stuck there. 
This means that all the kinetic energy  
gained by the moving charges in the absence of Ohmic losses, should get transferred to the plates of C$_2$, 
which we assume to be not free to move 
(clamped to the lab bench!). There will thus be necessarily inelastic collisions 
and the plates of C$_2$ would get heated because of these inelastic collisions. 
There could also be some partial energy loss in sparks but as we show later the {\em whole} energy loss cannot be accounted 
for by the radiation.

Actually $R\rightarrow 0$ is only a mathematical idealization which may not hold good 
when we go below certain very low resistance values. Let us take a material which can turn into a 
superconductor, say lead. If we lower its temperature, the resistance of the conductor will reduce steadily up to a 
certain point (7.22 K for lead),\cite{3} below which it may suddenly become zero as the material turns into a superconductor. That means 
either it will be a normal electrical resistance  with Ohmic losses above this turnover point or it will be zero 
resistance without Ohmic losses below this point. 
Thus there is a discontinuity in resistance and one does not have $R\rightarrow 0$ in limit.

Let us examine the idea of $R\rightarrow 0$ in limit in a non--superconductor material. 
Resistance of a wire is $R=\rho L/A$ where $\rho$ is the resistivity, $L$ is its  
length and $A$ is the cross section. We cannot increase $A$ beyond certain values (for example, it cannot be 
larger than the capacitor plate size), so we can decrease $\rho$ or/and $L$ to reduce $R$.  
Now $\rho \propto 1/\lambda$, the mean free path, meaning $R \propto L/\lambda$.
Usually $L/\lambda \sim 10^7$ for a few cm long wire, however starting from some finite resistance, as we
go to lower $R$, by decreasing $\rho$ and thereby increasing $\lambda$  
or decreasing $L$, the ratio $L/\lambda$ will decrease. And near some critical value of resistance, say $R_c$, 
$\lambda$ will approach $L$, that is the mean free path will become equal to the length of the wire or 
channel joining the two capacitors. At this stage $1/e$~th fraction of the current carrying 
charges will pass the length of the wire without suffering any collisions and thus without undergoing 
Ohmic losses. The remaining charges will of course undergo Ohmic losses due to collisions. 
It is of course statistically a random process. Let us denote 
the electric current by the latter as $I_1$ and that by the collisionless charges as $I_2$. Then $P_1=V I_1$ fraction 
will be the Ohmic losses and $P_2=V I_2$ fraction will be the power going into the kinetic energy of charges.
Thus there will be sharing of power losses between the two processes, with total power loss as 
$P=P_1+P_2=V (I_1+I_2)=V I$. 

Now let us see what will happen as we reduce $R$. Initially with much higher resistance than $R_c$, 
with $L/\lambda \gg 1$, there will be only $P=P_1$, the usual Ohmic losses. As we reach $R_c$, 
the Ohmic losses ($P_1$ fraction) will steadily decrease while the $P_2$ fraction will increase. 
For much lower resistance than $R_c$, there will be almost no collisions, there will be only $P=P_2$,
with the conducting charges gaining the kinetic energy in the absence of 
collisions and the $P_1$ losses being zero. 
\begin{figure}[ht]
\scalebox{0.45}{\includegraphics{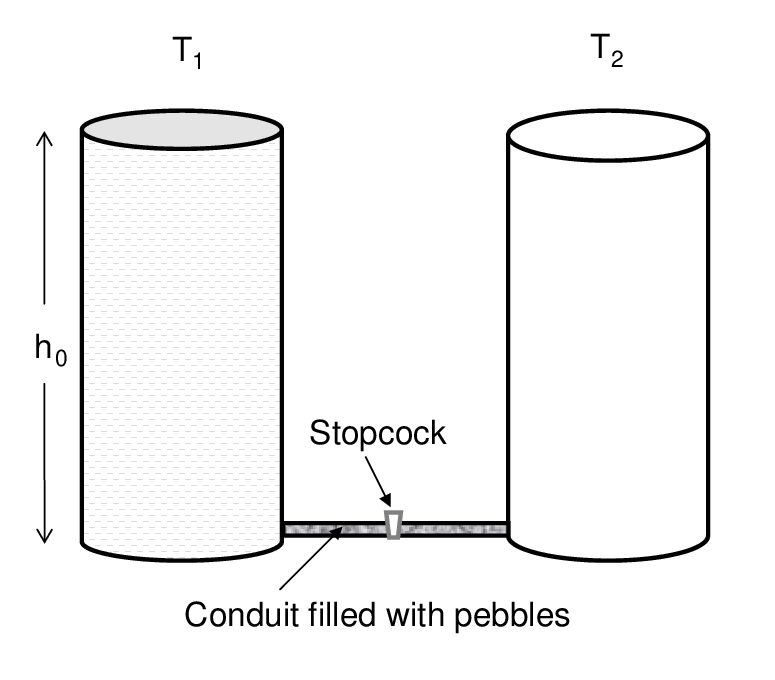}}
\caption{The equivalent case of missing-energy during transfer of water between two tanks of equal 
storage capacity }
\end{figure}

In (\ref{1}) it is implicitly assumed that all charges undergo Ohmic losses however low 
the collision rates might be (even when $R\rightarrow 0$), and accordingly the dissipation losses are calculated. 
In reality it may not even be proper to still think of resistance below $R_c$ in the usual ohm's law sense, when the 
collisions will be few and far between. Therefore $R\rightarrow 0$ might not be very meaningful much below $R_c$.
Thus the mysterious difference between $R=0$ and $R\rightarrow 0$ cases appears only because in the latter it is 
implicitly assumed that the charges lose their kinetic energy into Ohmic losses however low 
their collision rates might be, and accordingly we calculate the dissipation losses in (\ref{1}), 
while in an identically zero resistance case, Ohmic losses are not even considered. 

\section{The equivalent case of water transfer between two tanks}
An equivalent example exists in case of a water transfer from one full tank to an identical empty 
tank under the force of gravity (Fig.~2)\cite{19,20}. Initially the gravitational potential energy of the water 
to a height $h_{0}$ in tank T${_1}$ is $U_0=\int^{h_{0}}_{0}\rho g A z  {\rm d}z = \rho g A h_{0}^2 /2=Q_{0}V_{0}/2$, 
where $\rho$ is the density of the water, $A$ is the cross-section area of each tank, $g$ is the acceleration due to 
gravity and $z$ is the vertical distance. Then 
$Q_{0}=\rho A h_{0}$ is the total quantity (mass) of water and $V_{0}=gh_{0}$ is the gravitational potential. Now we 
open the stopcock, so that water is transferred from tank T${_1}$ into tank T${_2}$ through a conduit (Fig.~2). 
However when we consider the friction  with the conduit walls and obstructions within (say, pebbles inside 
the conduit blocking a free flow of water) then the water loses all its kinetic energy during the transfer 
to tank T${_2}$. At the end with each tank having $Q_{0}/2$ amount of water up to height $h_{0}/2$, the potential 
energy of the water in each tank is $Q_{0}V_{0}/8=U_0/4$ with the total energy of the system being $U_0/2$, 
exactly as in the two capacitor case. This is because the water in the upper half of tank T${_1}$ goes  
into the lower half of tank T${_2}$, then half of the total water mass (i.e., $Q_{0}/2$) which earlier was at a 
height between $h_0/2$ and $h_0$ in T${_1}$ is now at a height between $0$ and $h_0/2$ in T${_2}$, thus ending up 
at an average height lower by $h_0/2$, implying an energy loss of $U_0/2$. 
If there is no friction with the conduit walls (and no obstructions 
within either), from Bernoulli's theorem\cite{1} (or from simple energy conversion between potential and kinetic 
energy) the water would exit with a velocity $v=\sqrt{2g(h_1-h_2)}$ or $\rho v^2/2=\rho \Delta V$, 
at any moment when the heights of water columns in T${_1}$ and T${_2}$ are $h_1$ and $h_2$ respectively, 
with a gravitational potential difference $\Delta V=g(h_1-h_2)$.    
The water will thus move in the conduit with a kinetic energy that could be even utilized with a suitable 
device attached to the conduit (a tiny electric power generator!) otherwise this energy will be carried to
the tank T${_2}$ and ultimately lost as heat there by the time things have settled down.

For unequal tank capacities, let T${_1}$ and T${_2}$ have cross-section $A_1$ and $A_2$ respectively. Then 
the total water that will get transfered from T${_1}$ to T${_2}$ is $Q_2=Q_0 A_2/(A_1+A_2)$, and the height of 
water columns in the two tanks will be $h=h_0 A_1/(A_1+A_2)$. That means this much amount of water 
would have fallen from a height of initial average value $(h+h_0)/2$ in T${_1}$ to a final average value $h/2 $ in T${_2}$, 
implying an  average height loss of $h_0/2$ and the loss in potential energy of $Q_2\: g h_0/2=Q_2 V_0/2=U_0 A_2/(A_1+A_2)$. 
It also shows readily why the loss of energy in the tank (charged capacitor) system is the total transferred water (charge) 
$Q_2$ multiplied by half of the initial potential, i.e., $V_0/2$. 

\section{Possibility of radiation losses}
In the radiation hypothesis the authors in general assume that the power losses (irrespective of the 
expression for radiation losses, (see e.g.,  (8), (9) and (10) in \cite{8}) can be 
written as $V_{X} I= P_{rad}$ and have thus put $V_{X} = V_{12}$,
where $V_{12}$ is the potential difference between the two capacitors. Thus their assumption directly 
leads to $P_{rad} = V_{12} I$ and therefore $\int P_{rad}\:{\rm d}t = \int V_{12} I\:{\rm d}t=\int V_{12} \:{\rm d}Q$. 
From our (\ref{2}) we know the right hand side is $CV_{0}^{2}/4$ irrespective of the time dependence of $V$.
No wonder authors also get $\int P_{rad}\:{\rm d}t = CV_{0}^{2}/4$, as that is a built--in assumption. 
This way one is {\em bound to get} the same final result of energy losses irrespective of any other 
details of the exact radiation process that might have been assumed 
(whether it is a magnetic dipole radiation like the authors\cite{8} assumed or some other process),
and which could therefore be chosen any arbitrary function of time. In this particular case the authors 
emphasize that charging/discharging is not instantaneous. But according to this procedure for any arbitrary $P(t)$ 
one could define radiation resistance as  $R_r=P(t)/I^2$, and then writing $V_{12}=I R_r$, one gets $P=V_{12}I$ 
which no wonder gives $\int P_{rad}\:{\rm d}t=CV_{0}^{2}/4$,
and actually that way one does not really prove anything about the radiation process.
It is not the radiation hypothesis that gets confirmed this way, it is only the a priori assumption 
of equating radiation losses $P_{rad}$ (or losses in any other way!) to $V_{12} I$ which begets the 
apparently right answer. For this one does not even need to derive any complicated formulae for 
radiation expressions and it does not prove in any way that the radiation is that of 
magnetic dipole or some other ``multipole''. Different assumption about the radiation process 
(whether it is electric dipole or magnetic dipole or some other multipole) only at most may give a 
different time dependence of function $V(t)$ or $Q(t)$, but as the time integral of 
total charge transferred will be $Q_{0}/2$ and voltage $V_{0}/2$, one is bound to get the 
result for {\em energy dissipated} as $CV_{0}^{2}/4=U_0/2$. 
Moreover when charging/discharging is not instantaneous, the Ohmic resistance is not identically zero and 
the lost energy $CV_{0}^{2}/4$ should then be distributed between dissipation 
in $R$ and radiation. But we find that the energy dissipation is fully satisfied
by the Ohmic losses alone (Eq.~(\ref{1})) even when the resistance reduces in limit to zero ($R\rightarrow 0$) 
and the radiation hypothesis is not at all needed. 

It is possible to estimate how much maximum radiation losses can be there. 
From Larmor's formula\cite{17} we know that the energy radiated by a non-relativistic charge 
accelerated for a time interval $\Delta t$ (and thus having gained a velocity $v=a\Delta t$ in the 
absence of Ohmic losses) is $2q^2 a^2\Delta t/3c^3$. For all the energy gained by the charge due to 
the potential difference to go into radiation implies  
\begin{equation}
\label{3}
qV_{12}=\frac{m v^2}{2}=\frac{2q^2 a^2\Delta t}{3c^3}=\frac{2q^2 a v}{3c^3}
\end{equation}
or 
\begin{equation}
\label{4}
\frac{v}{a}=\Delta t=\frac{4q^2}{3m c^3}\sim\frac{r_e}{c},
\end{equation}
where $r_e=q^2/m c^2$ is the classical electron radius.\cite{17} Thus for {\em all} of the missing energy $U_0/2$ 
in the capacitor paradox to appear as radiation is possible {\em if and only if} the charges move from 
one capacitor to the other in a time interval of the order in which light travels the classical 
radius of the electron $r_e$, which is an impossible condition. In fact the radiation losses, 
due to the acceleration of the charges will be extremely small and can be made arbitrarily 
small by making the time over which the charge moves from $C_1$ to $C_2$ large enough.
For example, an external agency using 
some electrical probe (``magic tweezers''),\cite{1} could pick up charges one by one from C$_1$ at 
a higher potential and deliver them to C$_2$ at a lower potential at a leisurely rate (quasi-statically) 
and the difference in the potential energy of these  
charges can be utilized by the transferring agency. There will be no radiation losses, nor will there 
be any Ohmic loses. We shall further discuss one such alternate 
example in the next section.

\section{A capacitor is charged without using resistive wires}
Instead of charging a capacitor C$_2$ from C$_1$ using a wire of zero resistance, we could 
pose the problem in a different way. Let us suppose that we can expand or stretch the plates of a 
capacitor quasi-statically so that each plate area becomes double of 
its previous value, but without changing the plate separation. 
For simplicity we assume a parallel plate capacitor with dimensions $a$ and $b$ of 
the capacitor plates much larger than the plate separation, $h$, 
so that the electric fields within the capacitor can be considered, with
negligible errors, to be uniform as in the case of infinite plates. 
Let $\sigma_0=Q_0/A$ be the initial uniform surface charge density on the two
oppositely charged plates, with $A=ab$ as the surface area of each plate. 
Then the electrostatic field is a constant, $4\pi \sigma_0$, 
in the region between the two plates which thus have a potential difference $V=4\pi \sigma_0 h$. 
The field of course is zero everywhere outside. 
The mutual force of attraction on each plate is 
$2\pi\sigma_0^{2}$ per unit area, and the electric potential 
energy $U_{0}$ accumulated in separating the two plates by a 
distance $h$ is $2\pi\sigma_0^{2} A\, h$. 
The capacity of a parallel plate capacitor is given by $C=A/(4\pi h)$,\cite{3} and
with energy $U_{0}=CV{^2}/2=Q{^2}/2C$.

With an expansion of the capacitor plates' areas by a factor of two, the charge density becomes half  
with the charges now distributed over its double charge capacity.
The final energy of the capacitor is now only half of the previous value and the problem returns to 
the standard two capacitor paradox.
The question again rises where has half of the energy gone. Now that there are no connecting wires 
with their resistance coming into picture, so we do not have to worry about Ohmic losses. There are no radiation losses 
either.
As it is a quasi-static expansion there is no gain in the kinetic energy of current carrier charges. 
But we still have a problem of the missing energy. 

Actually in addition to the force of attraction between two plates of a capacitor, there is also an outward force of 
repulsion within each capacitor plate. 
The presence of such self-repulsive forces within the capacitor plates and the work done against them during a 
Lorentz contraction of the system when the charged capacitor system moves from one inertial frame to another, 
was first shown explicitly by Singal\cite{4} and accordingly the famous Trouton-Nobel experiment\cite{2} 
was resolved from energetic points of view.\cite{5} Here we will show by explicit calculations that 
the energy spent by the capacitor system during expansion is indeed equal to the missing energy, 
i.e, $CV_0^2/4$.  

Adapting the calculations of \cite{4} to our present case, we have calculated these  
force of self-repulsion in Appendix, where we find the expression for the rate of work done during an expansion 
of capacitor plates by the forces of self-expulsion as ${\rm d}{\cal W}/{\rm d}\eta=U_{0}\,/\eta^{2}$ 
(c.f. (\ref{26})) with $\eta$ as the expansion factor.

Now integrating from initial $\eta=1$ to a final expansion factor $\eta_{0}$, we get the amount of 
work done by the system during an expansion as ${\cal W}=U_{0}\,\left[1-1/\eta_{0}\right]$, which 
is equivalent to the energy loss $\Delta U$ in (\ref{1a}) with the charge capacitance having increased by a factor 
$\eta=(C_1+C_2)/C_1$. In particular, 
for $\eta_{0}=2$, we get the work done during expansion as ${\cal W}=U_{0}/2=Q_0^2/4C=CV_{0}^2/4$, 
which indeed is the energy that were missing in the two equal--capacitor problem. 

The above expression for energy change of the capacitor is quite general and it
shows that if $\eta_{0}\rightarrow \infty$, whole of the capacitor energy goes 
into the expansion of the plates (again this amounts to loss of all stored energy in (\ref{1a}) for  
$C_2\rightarrow \infty$). We can look at it in another way. If we were to contract 
the system ($\eta_{0}<1$), then we (an external agency!) have to do work against the forces of 
electrical self-repulsion within the capacitor plates. In fact the energy stored in the capacitor 
is nothing but the work done in bringing 
the charged capacitor plates from an infinite size to finite dimensions which is essentially the work 
done in moving the charges from infinity (against their electrical forces of mutual repulsion) 
to the finite-sized plates of a capacitor.

\section{Conclusion}
We have shown that the famous paradox of two charged capacitors is successfully resolved 
if one properly considers all the energy changes in the system. It was shown that the ``missing energy'' 
goes into the kinetic energy of conducting charges when the connecting wire has 
an identically zero resistance. 
The problem was formulated in an alternate form, without involving connecting wires in a circuit, 
where the capacitance of the system is increased by stretching the plates of the original capacitor. 
The paradox was properly resolved by showing that the work done by the outward self-forces, arising 
due to mutual repulsion among charges stored within each capacitor plate, during an expansion 
is equal to the missing energy of the capacitor system.
It was also shown that radiation plays no significant role in resolving the paradox. 
\section*{Acknowledgments}
I first learnt of this intriguing paradox from a talk by Prof. S. C. Dutta Roy of IIT Delhi 
in a conference where he exhorted the audience for a successful solution of this 
yet unresolved problem of many years, and where he also distributed hard copies of the 
transparencies of his talk to the interested people. 
\section{Appendix}
\subsection {Work done during a stretching of the plates of an ideal capacitor}
\begin{figure}[ht]
\scalebox{0.4}{\includegraphics{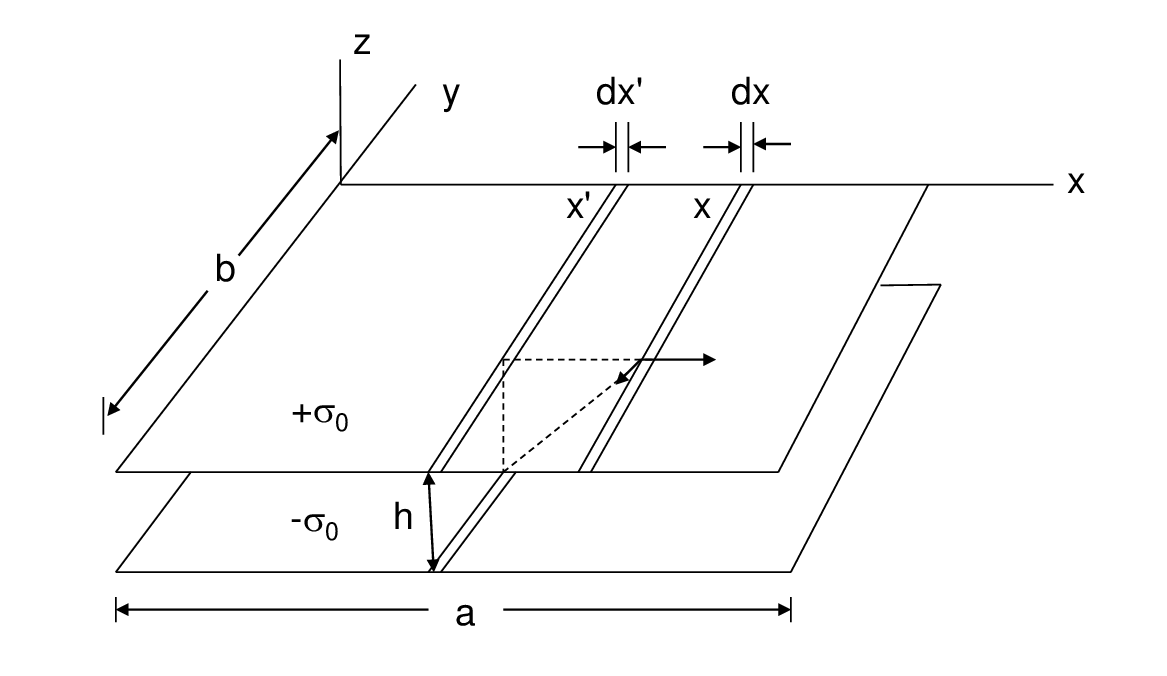}}
\caption{The geometry of the parallel plate capacitor for calculating the forces of 
self-repulsion within each plate of the capacitor. $\sigma_0$ is the surface charge density.}
\end{figure}

By an ideal capacitor we mean here that the surface charge density is uniform throughout on both 
plates. We assume that the charges somehow remain ''glued'' on the surface and the 
surface charge density decreases as the rubber--like plate surfaces are stretched. 
Let us assume the plates to be lying in the $x$-$y$ plane (Fig.~3). The 
electric field between the plates is parallel to the $z$-direction. 
The potential energy of the system as well as the energy in the electrostatic field is
$U_{0}=2\pi\sigma_0^{2}abh$, where $a,b$ are the plate dimensions and $h$ is the plate separation. 

Let us assume that we expand the plate dimensions by say, stretching them along the $x$-axis.
It should be noted that there are electromagnetic forces of {\em repulsion}
on charges {\em within} each plate, along its surface. We may generally ignore 
these repulsive forces, but during a stretching of the plates  parallel to the 
plate surface, work will be done by these forces. The forces
are indeed small near the plate-centers and become appreciable as we go away from
the plate centers, becoming maximum near the plate-edges, and it might seem that 
for $a$ and $b$ large enough as compared to $h$, the effect of these forces
should be negligible. But as we will see below, the amount of work done 
by theses forces during a plate expansion is proportional to the plate dimensions.

As the expansion considered is along the $x$-axis alone, then only the $x$-component
of the forces of repulsion will be relevant for our purpose. Now
the mutual electrostatic force of repulsion between two line charges,
each with a linear charge density $\lambda$ and of a length $b$,
separated by a distance $x$ is easily calculated to be
$2\lambda ^{2}(\sqrt{b^{2}+x^{2}}-x)/x$. 

\begin{figure}[ht]
\scalebox{0.4}{\includegraphics{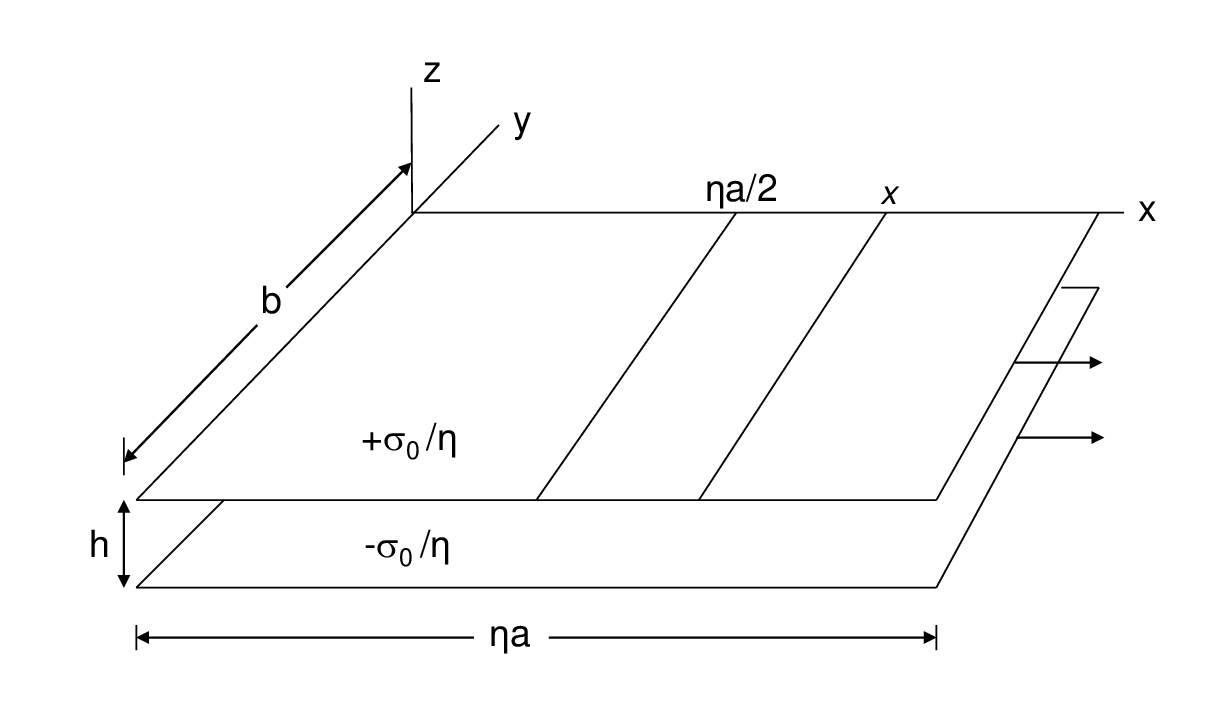}}
\caption{The parallel plate capacitor expanded by a factor $\eta$ with $\sigma_0/\eta$ 
as the surface charge density.}
\end{figure}
\begin{widetext}
Accordingly the net force of repulsion on a line charge of linear charge 
density $\lambda=\sigma_0 \,{\rm d}x$ lying at $x$, due to both plates is given by, 
\begin{eqnarray}
\label{16}
\nonumber
{\cal F}{\rm d}x \;=\; 2\sigma_0^2\,{\rm d}x\left[\int_{0}^{2x-a}{\rm d}x'\,\frac{\sqrt{b^2+(x-x')^2}-(x-x')}{x-x'}
-\int_{0}^{2x-a}{\rm d}x'\,\frac{x-x'}{\sqrt{h^{2}+(x-x')^{2}}}\right.\\
\cdot \left.\frac{\sqrt{b^{2}+h^{2}+(x-x')^{2}}-\sqrt{h^{2}+(x-x')^{2}}}{\sqrt{h^{2}+(x-x')^{2}}}\right]. 
\end{eqnarray}

Here the second integral term represents the $x$-component
of the force of attraction on the line element at $x$ 
due to the oppositely charged plate lying
at a distance $h$ below (Fig.~3). We have taken the line element at $x$
to be in the right-half of the plate, which experiences a net force
towards the +ve $x$-axis; the left-half of each plate would equally
experience a net force along the $-$ve $x$-axis. Further, only the
portion of each plate lying between 0 and $2x\!-\!a$ contributes a net 
force at $x$, the force due to the remaining portion of each plate 
gets cancelled because of its symmetry about $x$.

With a change of variable $x-x'=\xi$, we can write,
\begin{equation}
\label{17}
{\cal F} =2\sigma_0 ^{2}\int_{a-x}^{x}{\rm d}\xi\:\:g(\xi)
\end{equation}
where
\begin{eqnarray}
\label{18}
g(\xi)=\frac{\sqrt{b^{2}+\xi ^{2}}-\xi }{\xi}-\frac{\xi }{h^{2}+\xi ^{2}}
\cdot\left(\sqrt{b^{2}+h^{2}+\xi ^{2}}-\sqrt{h^{2}+\xi ^{2}}\right).
\end{eqnarray}
Fig.~4 shows a capacitor whose plates have undergone a uniform expansion by a factor 
$\eta$ and accordingly the charge density reduced to $\sigma_0 /\eta$. Now the charges  
at $x$ on the expanded plates move an infinitesimal distance $(x-\eta a/2){\rm d}\eta/\eta$ 
further away with respect to the plate centers, during the change in expansion factor from 
$\eta$ to $\eta+{\rm d}\eta$. 

Then the rate of work being done by the forces of self-interaction, during expansion 
of {\em both plates}, is written as ,
\begin{equation}
\label{19}
{\rm d}{\cal W}=8\frac{\sigma_0 ^{2}}{\eta^{3}}{{\rm d}\eta}\int_{\eta a/2}^{\eta a}{\rm d}x
\:(x-\eta a/2)\int_{\eta a-x}^{x}{\rm d}\xi\:\:g(\xi).
\end{equation}
One factor of 2 in the above expression has entered because an equal work is done on 
both halves of either plate, while another factor of 2 arose because work is done 
during expansion of each of the plates.\\
The rate of work done can be written as
\begin{equation}
\label{20}
{\rm d}{\cal W}=4\frac{\sigma_0 ^{2}}{\eta^{3}}{{\rm d}\eta}
\int_{0}^{\eta a}(2x-\eta a)f(x)\:{\rm d}x,
\end{equation}
where 
\begin{eqnarray}
\nonumber
\label{21}
f(x) =\int{\rm d}\xi\:\:g(\xi) = \sqrt{x^{2}+b^{2}}-x +\sqrt{x^{2}+h^{2}}-\sqrt{x^{2}+b^{2}+h^{2}}\\
 -\: b\,\ln\left(\frac{\sqrt{x^{2}+b^{2}+h^{2}}-b}{\sqrt{x^{2}+h^{2}}}\cdot\frac{x}{\sqrt{x^{2}+b^{2}}-b}\right).
\end{eqnarray}
With the help of the indefinite integrals,
\begin{eqnarray}
\label{22}
\nonumber
\int\ln\left(\frac{\sqrt{x^{2}+b^{2}+h^{2}}-b}{\sqrt{x^{2}+h^{2}}}
\right){\rm d}x= x\ln\left(\frac{\sqrt{x^{2}+b^{2}+h^{2}}-b}{\sqrt{x^{2}+h^{2}}}\right)+ b\ln\left(\sqrt{x^{2}+b^{2}+h^{2}}-x\right)\\
+h\tan^{-1}\, \frac{bx}{h\,\sqrt{x^{2}+b^{2}+h^{2}}},
\end{eqnarray}
and 
\begin{eqnarray}
\nonumber
\label{23}
\int x\,\ln\left(\frac{\sqrt{x^{2}+b^{2}+h^{2}}-b}{\sqrt{x^{2}+h^{2}}}
\right){\rm d}x  =  \frac{1}{2}(x^{2}+h^{2})
.\ln\left(\frac{\sqrt{x^{2}+b^{2}+h^{2}}-b}{\sqrt{x^{2}+h^{2}}}\right)\\
 -  \frac{b}{2}\,\sqrt{x^{2}+b^{2}+h^{2}},
\end{eqnarray}
and after a simplification, we finally get the following 
expression for the rate of work done during an expansion of the system,
\begin{eqnarray}
\nonumber
\label{24}
{\rm d}{\cal W}=4\sigma_0^{2}
\left[\frac{2h^{2}}{3}\left(\sqrt{\eta^{2}a^{2}+h^{2}}-h+\sqrt{b^{2}+h^{2}}-\sqrt{\eta^{2}a^{2}+b^{2}+h^{2}}\right)\right.\\
\nonumber
-\frac{b^{2}}{3}\left(\sqrt{b^{2}+h^{2}}-b+\sqrt{\eta^{2}a^{2}+b^{2}}-\sqrt{\eta^{2}a^{2}+b^{2}+h^{2}}\right) \\
\nonumber
+\frac{\eta^{2}a^{2}}{6}\left(\sqrt{\eta^{2}a^{2}+h^{2}}-\eta a+\sqrt{\eta^{2}a^{2}+b^{2}}-\sqrt{\eta^{2}a^{2}+b^{2}+h^{2}}\right)\\
\nonumber
+\frac{\eta ab^{2}}{2}\,\ln\left(\frac{\sqrt{\eta^{2}a^{2}+b^{2}+h^{2}}-\eta a}
{\sqrt{b^{2}+h^{2}}}\cdot\frac{b}{\sqrt{\eta^{2}a^{2}+b^{2}}-\eta a}\right)\\
\nonumber
-\frac{\eta ah^{2}}{2}\,\ln\left(\frac{\sqrt{\eta^{2}a^{2}+b^{2}+h^{2}}-\eta a}
{\sqrt{b^{2}+h^{2}}}\cdot\frac{h}{\sqrt{\eta^{2}a^{2}+h^{2}}-\eta a}\right)\\
\nonumber
-bh^{2}\,\ln\left(\frac{\sqrt{\eta^{2}a^{2}+b^{2}+h^{2}}-b}
{\sqrt{\eta^{2}a^{2}+h^{2}}}\cdot\frac{h}{\sqrt{b^{2}+h^{2}}-b}\right) \\
\left.+\,\eta abh\,\tan ^{-1}\, \frac{\eta ab}{h\,
\sqrt{\eta^{2}a^{2}+b^{2}+h^{2}}}\right]\frac{{\rm d}\eta}{\eta^{3}}. 
\end{eqnarray}
We can expand this complicated-looking expression in terms of an
ascending power series in $h/\eta a$ , $h/b$ , $h/\sqrt{\eta^{2}a^{2}+b^{2}}$  as
\begin{equation}
\label{25}
{\rm d}{\cal W}=4\sigma_0 ^{2}\,\eta abh\left[\frac{\pi}{2}+O
(\frac{h}{\eta a},\frac{h}{b},\frac{h}{\sqrt{\eta^{2}a^{2}+b^{2}}})\right]
\frac{{\rm d}\eta}{\eta^{3}},
\end{equation}
where $O(\cdots )$ represents the first and higher order power series terms in $h/\eta a , h/b$ etc. 

Therefore for $h\ll \eta a,b,$ we get, 
\begin{equation}
\label{26}
\frac{{\rm d}{\cal W}}{{\rm d}\eta}=\frac{2\pi\sigma_0^2 abh}{\eta^{2}}=\frac{U_0}{\eta^2}.
\end{equation}
\end{widetext}

\begin{thebibliography}{33}
\expandafter\ifx\csname natexlab\endcsname\relax\def\natexlab#1{#1}\fi
\expandafter\ifx\csname url\endcsname\relax
  \def\url#1{{\tt #1}}\fi
\expandafter\ifx\csname urlprefix\endcsname\relax\def\urlprefix{URL }\fi
\bibitem{1}D. Halliday, R. Resnick and J. Walker, {\em Fundamentals of Physics}, 5th ed. 
John Wiley, New Jersey 1997, Ch. 15, 26, 27


\bibitem{6}R. A. Powel, 
 Am. J. Phys. {47}, (1979) 460-462


\bibitem{7}K. Mita and M. Boufaida,  Am. J. Phys. {67}, (1999) 737-739 

                                  
\bibitem{10}C. Zucker, Am. J. Phys. {23} (1955) 469-469


\bibitem{11}R.P. Mayer, J.R. Jeffries and G.F. Paulik, IEEE Trans. Education {36} (1993) 307-309


\bibitem{12}K. Lee, Eur. J. Phys. {30} (2009) 69-74


\bibitem{13}C. E. Mungan, Eur. J. Phys. {30} (2009) L59-L63


\bibitem{14}A. M. Abu-Labdeh and S. M. Al-Jaber, J. Electrostatics {66} (2008) 190-192


\bibitem{15}W. J. O'Connor, Phys. Educ. {32} (1997) 88


\bibitem{16}S. Mould, Phys. Educ. {33} (1998) 323 


\bibitem{18}A. M. Sommariva, IEE Proceedings - Circuits, Devices and Systems  150, issue 3 (2003) 227-231


\bibitem{19}D. P. Korfiatis, WSEAS Trans. Circuits and Systems {6} (2007) 76-79


\bibitem{8}T. B. Boykin, D. Hite and N. Singh, Am. J. Phys. {70} (2002) 415-420


\bibitem{9}T. C. Choy, Am. J. Phys. {72} (2004) 662-670


\bibitem{3}E. M. Purcell,  Electricity and Magnetism, 2nd ed., McGraw-Hill, New York 1985, Ch. 3, 4, Appendix C


\bibitem{20}S. Krishnan and M. Rao, \newblock Am. J. Phys. {50} (1982), 662


\bibitem{17}J. D. Jackson, \newblock {Classical Electrodynamics, 2nd ed., John Wiley,  New York 1975, Ch. 14}


\bibitem{4}A. K. Singal,  \newblock J. Phys. A {25} (1992), 1605--1620


\bibitem{2}F. T. Trouton and  H. R. Noble, 
\newblock Phil. Trans. Roy. Soc. London A {202} (1903), 165-181


\bibitem{5}A. K. Singal, \newblock Am. J. Phys. {61} (1993), 428-433

\end{thebibliography}

\end{document}